# Dual-Mode Asymmetric Transmission based on Asymmetric and Orthogonal Gratings: Polarization-Dependent and -Independent Modes

Ruihan Ma[1], Yuqing Cheng[1,*] and Mengtao Sun[1,†]

[1] School of Mathematics and Physics, University of Science and Technology Beijing, Beijing 100083, People's Republic of China

**Abstract:** A dual-mode asymmetric transmission (AT) nanodevice based on the asymmetric and orthogonal grating-film-grating (AO-GFG) structure is proposed and systematically investigated theoretically. The device supports two distinct localized surface plasmon resonance (LSPR) modes for forward transmission, corresponding to the polarization-dependent ($M_1$) and the polarization-independent ($M_2$) resonances, respectively. This results in the fact that when x-polarized light is incident, only $M_2$ exists; when y-polarized light is incident, both $M_1$ and $M_2$ exist. Besides, both modes yield the maximum isolation ratio of more than 10 dB. The electric field distributions further indicate that $M_2$ exhibits strong confinement and efficient tunneling through the metallic film, while $M_1$ shows weaker but more polarization-sensitive hybridization. The coexistence and tunability of these two modes constitute the physical basis of dual-mode AT, highlighting the AO-GFG structure as a promising platform for high-isolation and polarization-tunable plasmonic devices in the visible and near-infrared regions.

## 1. Introduction

Asymmetric transmission (AT) refers to the unequal transmission responses of an optical system for forward and backward transmission under a given polarization state. In recent years, the realization of AT has attracted significant interest in polarization-sensitive photonic devices[1-3], optical information encoding[4, 5], secure communication[6, 7], and on-chip photonic systems[8-11]. To achieve AT responses, various physical mechanisms and structural designs have been proposed, including chiral metamaterials[12-15], anisotropic metasurfaces[16-20], multilayer dielectric gratings[21-24], and plasmonic nanostructures[25-28]. Among these approaches, asymmetric gratings have emerged as a highly promising platform owing to their structural simplicity, ease of fabrication, and superior performance in manipulating direction-dependent transmission. By breaking the spatial symmetry of periodic gratings[29-32], AT gratings enable efficient control of light propagation, providing a versatile pathway for integrating asymmetric transmission into compact photonic circuits. Luan et al. developed a dual-wavelength metadevice using an ITO-based modular design, operating at 2365 nm in reflection and 650 nm in transmission, and demonstrated flexible beam deflection and focusing with minimal crosstalk[33]. Battal et al. designed a subwavelength-slit device with double-layer gratings and a dielectric spacer, achieving an ultrahigh contrast ratio of 53.4 dB at 1.56 μm with strong unidirectional transmission[34]. Cheng et al. previously introduced a hybrid metallic nanowaveguide, in which asymmetric gratings were incorporated at both the input and output ports to realize asymmetric transmission with multiple modes in the visible spectrum[35].

[*] Email: yuqingcheng@ustb.edu.cn
[†] Email: mengtaosun@ustb.edu.cn

In this study, we employ the asymmetric and orthogonal grating-film-grating (AO-GFG) structure to achieve dual-mode AT with one mode (around 789 nm) polarization-dependent and the other mode (around 814 nm) polarization-independent. The proposed structure offers a compact and feasible strategy for realizing polarization-tunable and high-isolation plasmonic transmission, providing new physical insight into multi-mode coupling mechanisms in asymmetric nanostructures.

## 2. Method and structure

Finite-difference time-domain (FDTD) method is employed to simulate the optical properties of the structure [16, 36]. The schematic of the AT nanodevice is shown in Fig. 1. It consists of a $SiO_2$ substrate, two orthogonal sets (one is arranged along the x-axis at x < 0 part of the unit cell, the other is along the y-axis at x > 0 part) of upper and lower Ag gratings, and a middle Ag film. Particularly, the parameters of the upper and lower gratings are different, leading to AT in transmissivity [37, 38]. In the simulation, the refractive index of SiO$_2$ is 1.45, and the one of Ag is from the experimental data of Ref. [39]. The incident light is linearly polarized plane wave along z (backward) or −z (forward) direction, with x-polarized corresponding to an incident angle of 0° and y-polarized corresponding to 90°. Period boundary conditions in x and y directions are employed. The parameters are optimized based on the forward transmissivity and the transmissivity isolation ratio between forward and backward transmission (IR$_T$) to enhance device performance. Here, IR$_T$ in unit dB is defined as:

$$IR_T = 10 \times \log_{10}\left(\frac{T_F}{T_B}\right) \qquad (1)$$

where T$_F$ and T$_B$ stand for the transmittivities of forward and backward transmission, respectively. The structural parameters are shown in Table 1. The thickness of the middle Ag film is $d = 20$ nm for the device. The coordinate origin is located at the center of the xy plane, with the lower surface of the middle film defined as z = 0.

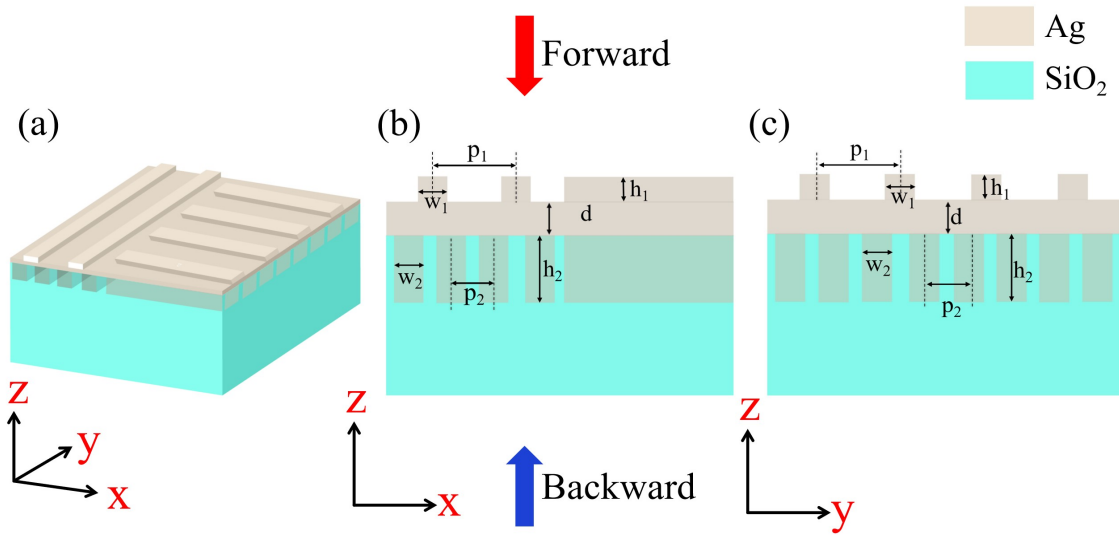

**Fig. 1.** Schematic of the AT device unit cell. (a) Three-dimensional view. (b) Front view (xz plane). (c) Side view (yz plane). The light gray represents the Ag ridges and film, the dark gray represents the Ag grooves, and the cyan represents SiO$_2$.

Table 1. Parameters and modes of the AT Device

| Parameters | unit: nm |
| --- | --- |
| Upper grating period ($p_1$) | 800 |
| Lower grating period ($p_2$) | 400 |
| Upper groove depth ($h_1$) | 30 |
| Lower groove depth ($h_2$) | 190 |
| Upper ridge width ($w_1$) | 270 |
| Lower ridge width ($w_2$) | 270 |

## 3. Results and discussion

The mechanism of the proposed AT device relies on the unidirectional excitation and tunneling of energy through localized surface plasmon resonance (LSPR). When polarized light is incident in the forward direction (along the −z axis), the upper grating provides the necessary wavevector to excite LSPR at the air/metal interface. The localized modes tunnel through a silver film with a thickness smaller than the LSPR penetration depth and are efficiently decoupled into the substrate by the lower grating, resulting in strong forward transmission.

Under backward incidence (along the z axis), the lower grating does not support wavevector conditions for LSPR excitation at the same frequency. Consequently, the incident light cannot generate LSPR and is mostly reflected by the silver film. Furthermore, near-field coupling between the upper and lower gratings suppresses backward transmission, leading to near-zero transmissivity in the backward transmission. Therefore, this kind of device can achieve a high isolation ratio exceeding 10 dB.

Overall, the device offers a promising platform for photonic systems exhibiting asymmetric transmission.

## 3.1 Asymmetric spectra

Figs. 2**(a)** and 2**(c)** respectively show the forward and backward transmission spectra of the device under x- (*Px*) and y-polarized (*Py*) incidence, respectively, while Figs. 2**(b)** and 2**(d)** illustrate the corresponding isolation ratios. These results are employed to systematically analyze the AT behavior and wavelength dependence of the structure.

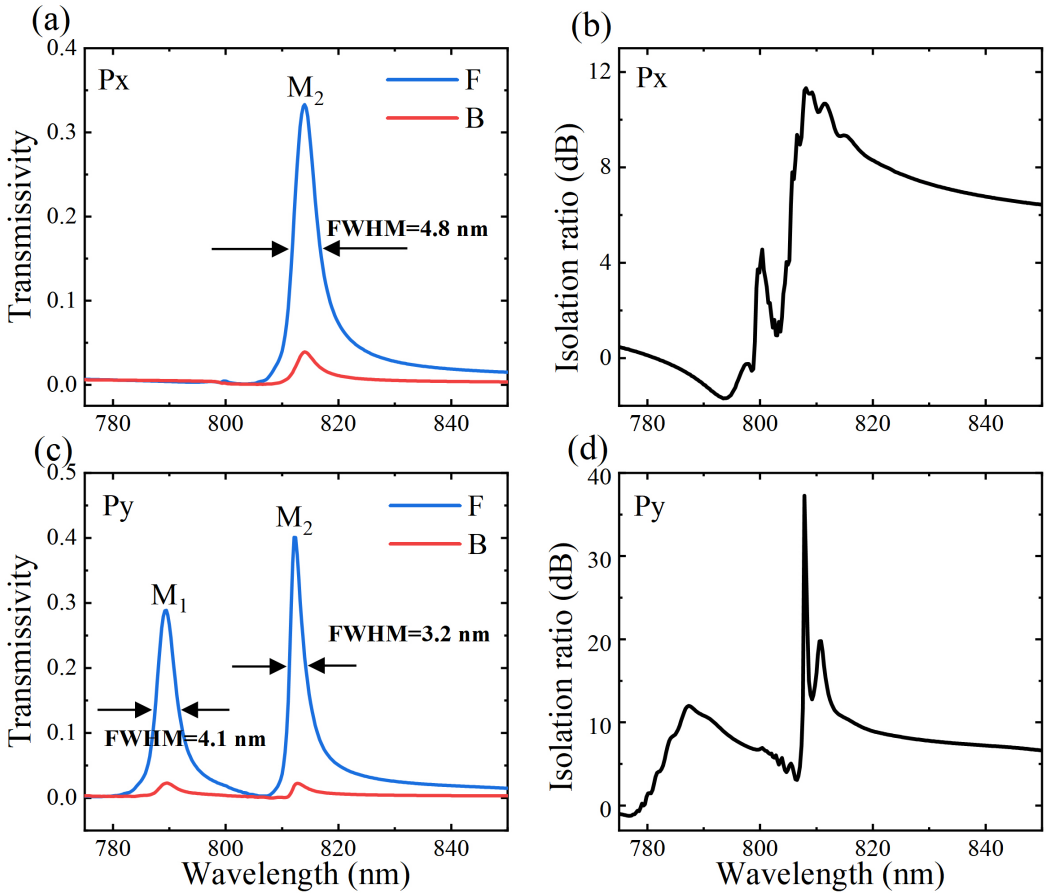

**Fig. 2.** The forward (F) and backward (B) transmission spectra with *Px* (a) and *Py* (c) polarized incidence. Blue and red curves stand for forward and backward transmissivities, respectively. Isolation ratio between forward and backward transmissivities for *Px* (b) and *Py* (d).

From the transmission spectra, it can be observed that at specific wavelengths, the forward transmissivity is significantly higher than the backward transmissivity, exhibiting a typical AT characteristic. This phenomenon is primarily attributed to the coupling effect between the upper and lower gratings and the metallic film, as well as the selective excitation of LSPR, whereby the polarization direction of the incident light governs the efficiency of the energy coupling pathways.

In terms of polarization dependence, there exists a pronounced distinction between the responses under x-polarization and y-polarization. Under x-polarization, a single sharp resonance corresponding to $M_2$ appears, as shown in Fig. 2**(a)**, where the forward transmissivity reaches approximately 0.33, while the backward transmissivity remains less than 0.05, indicating a strong suppression of backward-propagating light. The isolation ratio reaches a maximum of 9.3 dB. In contrast, under y-polarization, the transmission spectrum exhibits a pair of resonances corresponding to the $M_1$ and $M_2$ modes, as shown in Fig. 2**(c)**. At both resonances, the forward transmissivity is significantly higher than the backward one, with the latter remaining below 0.02 and the isolation ratios reaching up to 11 dB and 12.7 dB for M1 and M2, respectively.

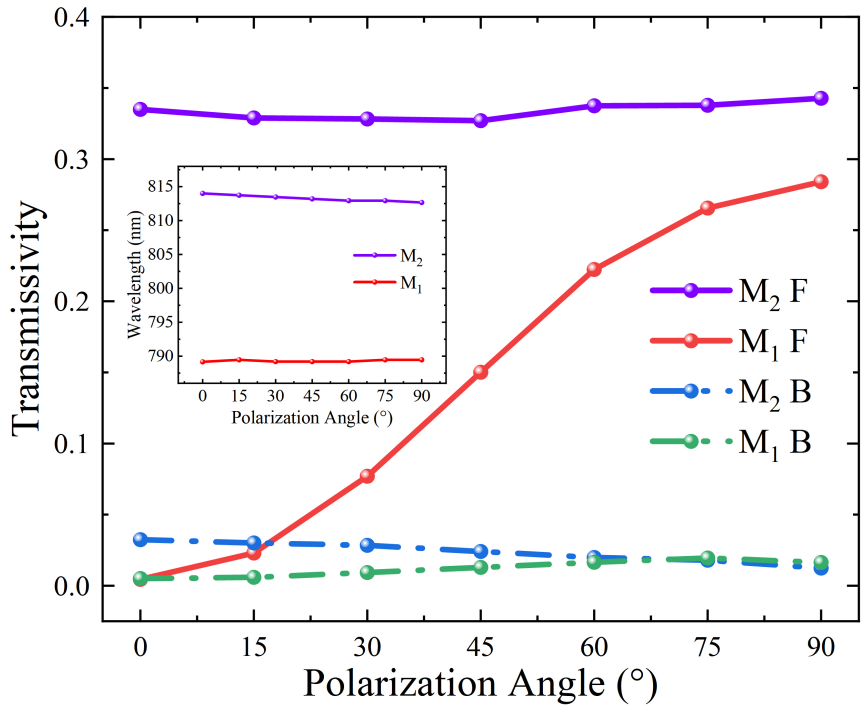

**Fig. 3.** Transmissivity of $M_1$ and $M_2$ as a function of the polarization angle of the incident light. Purple and red solid lines stand for forward transmission of $M_2$ and $M_1$, respectively. Blue and green dot lines stand for backward transmission of $M_2$ and $M_1$, respectively. The inset stands for the resonance wavelengths of $M_1$ and $M_2$ as a function of polarization angle.

Fig. 3 further shows the transmissivity of the two modes varying with the incident polarization angle. As the polarization angle increases from 0° to 90°, the forward transmissivity of $M_2$ remains essentially constant, while the one of $M_1$ monotonically increases. Meanwhile, the backward transmissivities of both modes are close to zero and is insensitive to the angle. Overall, the device exhibits robust asymmetric transmission for both polarizations, with $M_2$ representing a polarization-independent mode that maintains stable transmission characteristics under different polarization states, and $M_1$ representing a polarization-dependent mode that is highly sensitive to polarization of incident light.

These results indicate that the proposed structure enables effective control of unidirectional transmission by tuning polarization and wavelength, thereby offering a feasible route to achieving high-isolation AT. Such capability holds significant implications for the design of integrated plasmonic photonic devices.

*3.2 Electric field modes*

In order to understand the mechanisms of this AT phenomenon, the electric field is calculated and illustrated. Define $M = E/E_0$ as the ratio of the electric field intensity (E) to the one of the source ($E_0$). Figs. 4-6 show the distributions of the z-component ($E_z$) and the in-plane components ($E_x$ and $E_y$) of the electric field in the xz plane at y = 150 nm under x- and y-polarized illumination, where $E_x$ and $E_y$ denote the x- and y-components of the electric field, respectively.

As shown in Fig. 4, the $E_z$ distributions reveal the distinct excitation and propagation characteristics of the device at the two resonant wavelengths.

First, we discuss the case of x-polarized incidence. For forward incidence, the incident wavevector matches the periodicity of the upper grating, enabling efficient excitation of localized surface plasmon resonance (LSPR) modes at the metal-air interface. For $M_2$ (Fig. 4(a)), $E_z$ is strongly enhanced and localized near the surface of, indicating that the LSPR mode can effectively tunnel through the thin Ag film.

Assisted by the lower grating, the energy subsequently decouples into the dielectric substrate, thereby establishing a robust transmission channel and resulting in high forward transmissivity. For $M_1$ (Fig. 4(c)), $E_z$ intensity is much weaker than the one of $M_2$, suggesting that the tunneling across the film is much less, resulting in the almost zero transmissivity. For backward incidence (Figs. 4(e) and 4(g)), the mismatch of the wavevectors results in the weak LSPR, which leads to the limited tunneling and transmission.

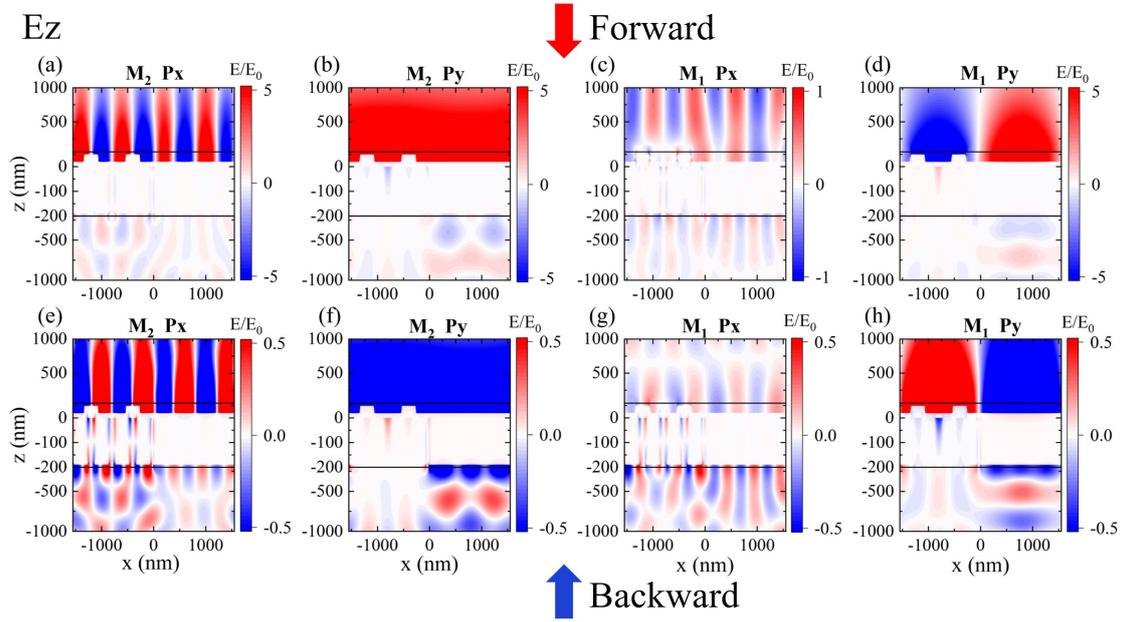

**Fig. 4.** The z-component of electric field distribution ($E_z$) in the xz plane for the two modes varying with the polarization at y = 150 nm plane. (a-d) Forward incidence: (a) $M_2$, *Px*. (b) $M_2$, *Py*. (c) $M_1$, *Px*. (d) $M_1$, *Py*. (e-h) Backward incidence: (e) $M_2$, *Px*. (f) $M_2$, *Py*. (g) $M_1$, *Px*. (h) $M_1$, *Py*. The color bar represents electric field intensity relative to the one of incident light.

Second, we discuss the case of y-polarized incidence. For forward incidence, both $M_1$ and $M_2$ illustrate strongly enhanced electric field ($E_z$), as shown in Fig. 4(b) and 4(d). Similar to the case of x-polarized incidence, it results in efficient tunneling through the Ag film and thus high transmissivities due to the decoupling of the lower gratings. For backward incidence as shown in Fig, 4(f) and 4(h), the weak LSPR mode couldn't support efficient tunneling thus low transmissivities.

These results confirm that the resonances at both modes correspond to strongly enhanced and localized LSPR mode for forward incidence, while the LSPR mode is not strong enough for backward incidence. The sharp contrast between forward and backward incidence provides a clear physical basis for the high isolation ratio achieved by the device.

To illustrate the polarization characteristics and coupling mechanisms of the dual-mode AT, the in-plane electric field components $E_x$ and $E_y$ are discussed under both forward and backward incidences, as shown in Figs. 5 and 6, respectively. Different from $E_z$, $E_x$ and $E_y$ are not localized but propagate to the free space.

For backward transmission, in all the cases, the in-plane electric intensities above plane z = 0 nm are all too weak to propagate to the far field, as shown in Figs. 5**(e-h)** and Figs. 6**(e-h)**. Therefore, the transmissivities of all the backward transmissions remain all almost zero. We will discuss the forward transmission in details below.

In the case of x-polarized incidence, the electric field propagating to the far field is principal the $E_x$ component, which can be illustrated by comparing Fig. 5**(a)** and 6**(a)**. Specifically, below plane z = −200 nm, only $M_2$ shows considerable $E_x$ intensity while $M_1$ shows very weak one, as shown in Fig. 5**(a)** and 5**(c)**, resulting in high and low transmissivities of $M_2$ and $M_1$, respectively. Besides, $E_x$ in Fig. 5**(a)** at both left (x < 0) and right (x > 0) sides shows considerable intensities, indicating that both the orthometric gratings contribute to the transmissivity of forward transmission with x-polarized incidence for $M_2$.

In the case of y-polarized incidence, the electric field propagating to the far field is principal the $E_y$ component, which can be illustrated by comparing Fig. 5**(b)** and 6**(b)**, or Fig. 5**(d)** and 6**(d)**. Specifically, below plane z = −200 nm, both $M_1$ and $M_2$ show considerable $E_y$ intensities, as shown in Fig. 6**(b)** and 6**(d)**, resulting in both high transmissivities of them. Besides, $E_y$ in Fig. 6**(b)** and 6**(d)** shows considerable intensities only at right (x > 0) side, indicating that only the right gratings contribute to the transmissivities of forward transmission with y-polarized incidence for the two modes. That is, the left gratings cannot support efficient forward transmissivity with y-polarized incidence.

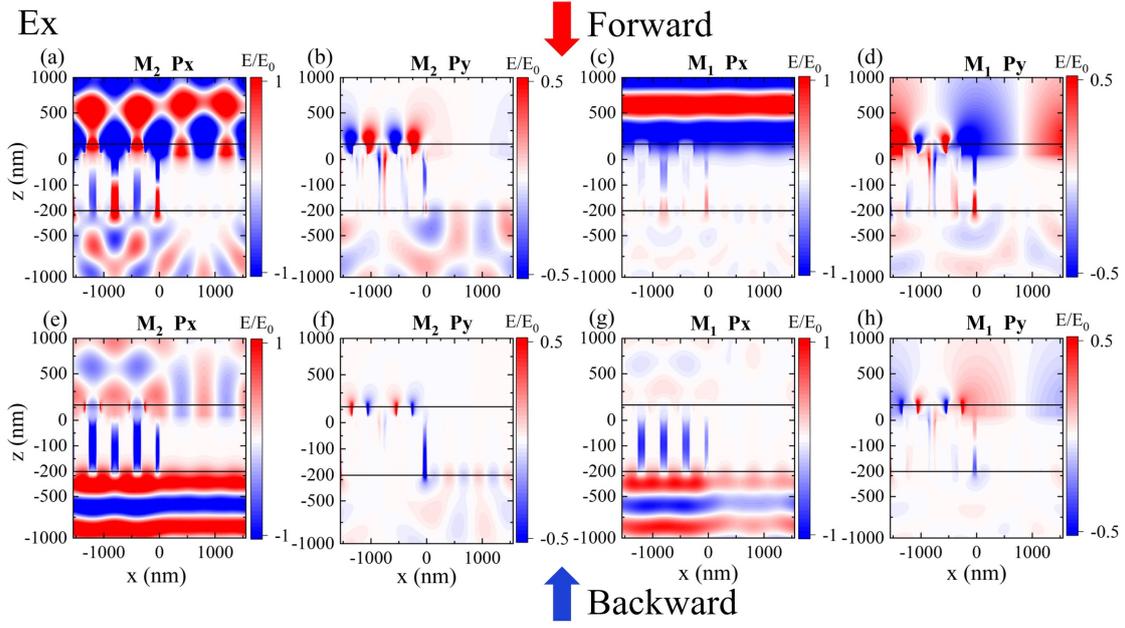

**Fig. 5.** Electric field distribution ($E_x$ component) in the xz plane for the two modes varying with the polarization. (a-d) Forward incidence: (a) $M_2$, *Px*. (b) $M_2$, *Py*. (c) $M_1$, *Px*. (d) $M_1$, *Py*. (e-h) Backward incidence: (e) $M_2$, *Px*. (f) $M_2$, *Py*. (g) $M_1$, *Px*. (h) $M_1$, *Py*. The color bar represents electric field intensity relative to the one of incident light.

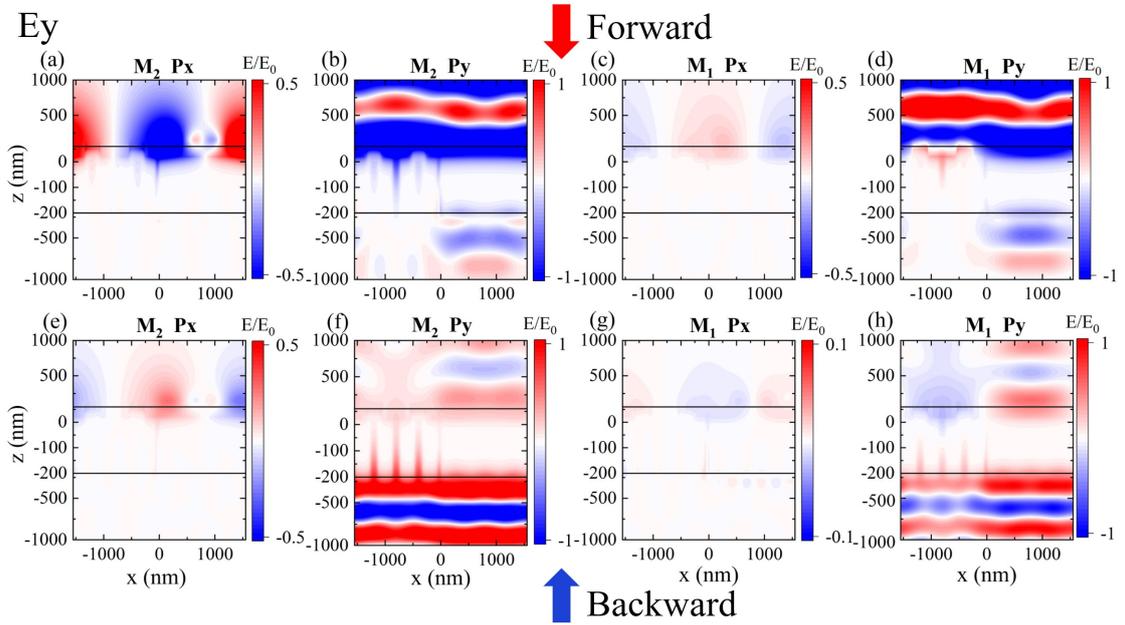

**Fig. 6.** Electric field distribution ($E_y$ component) in the xz plane for the two modes varying with the polarization. (a-d) Forward incidence: (a) $M_2$, *Px*. (b) $M_2$, *Py*. (c) $M_1$, *Px*. (d) $M_1$, *Py*. (e-h) Backward incidence: (e) $M_2$, *Px*. (f) $M_2$, *Py*. (g) $M_1$, *Px*. (h) $M_1$, *Py*. The color bar represents electric field intensity relative to the one of incident light.

From the above analysis, we conclude that the x < 0 gratings support efficient forward transmission with x-polarized incidence only for $M_2$, while the x >0 gratings support efficient forward transmission with x-polarized incidence also only for $M_2$, but with y-polarized incidence for both $M_2$ and $M_1$. This property of the structure results in the polarization-independent mode ($M_2$) and polarization-dependent mode ($M_1$).

Fig. 7 shows the total electric field distribution of the device in the xy plane for forward transmission. The total electric field is mainly dominated by the z component, while the contributions of the x and y components are relatively small (not shown), which is consistent with the results of $E_z$, $E_x$, and $E_y$ presented in Figs 4-6.

At the resonance of $M_2$ (Figs. 7**(a)**-**(b)** and 7**(e)**-**(f)**), the electric field is strongly enhanced and localized at the grooves of the upper surface, whereas the intensity at the lower surface is comparable to the one of source ($E_0$). In particular, with x-polarized incidence (Fig. 7**(a)**), the x < 0 gratings contribute to the transmissivities; while with y-polarized incidence (Fig. 7**(b)**), the x > 0 gratings contribute to the transmissivities. This indicates the mechanism of the polarization-independent property of $M_2$.

At the resonance of $M_1$ (Figs. 7**(c)**-7**(d)** and 7**(g)**-7**(h)**), the electric field is strongly enhanced and localized at the grooves of the upper surface only for y-polarized incidence. In particular, with x-polarized incidence, these gratings cannot support efficient transmissivities; while with y-polarized incidence, the x > 0 gratings can support efficient transmissivity and make the major contribution. This indicates the mechanism of the polarization-dependent property of $M_1$.

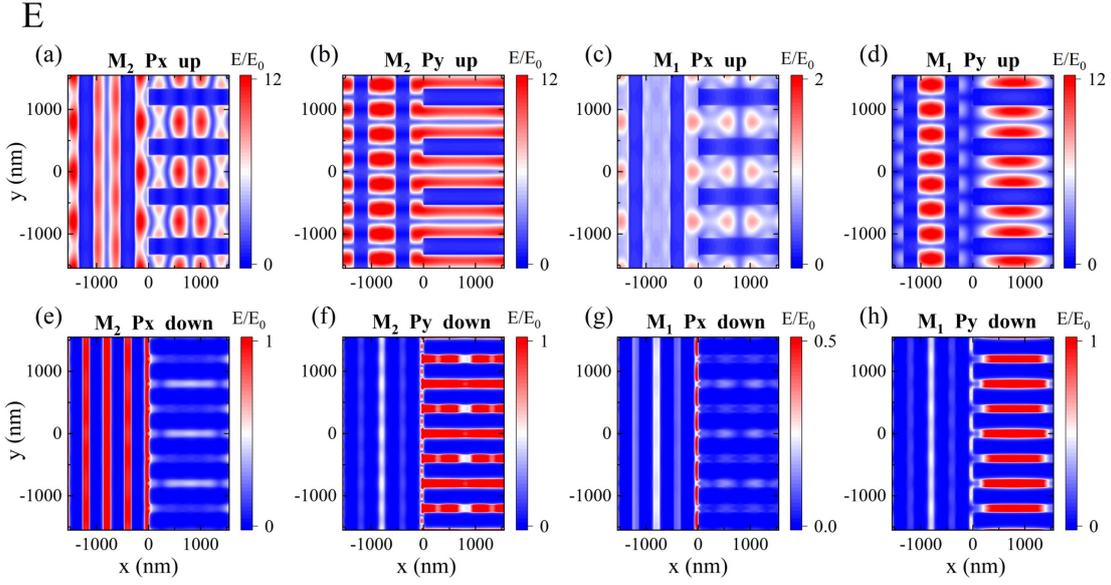

**Fig. 7.** Total electric field (E) distribution in the xy plane for forward transmission of the two modes varying with the polarization of forward incidence. The upper row (a-d) represents to the upper interface (up) z = 35 nm, and the lower row (e-h) represents to the lower interface (down) z = −95 nm. The color bar represents electric field intensity relative to the one of incident light.

The above analysis for Fig. 7 lies the same conclusion of the ones for Figs. 4-6 about the polarization-dependent and -independent mechanism of the two modes.

## 4. Conclusion

In summary, the AT behavior of the proposed AO-GFG nanostructure has been theoretically investigated and physically clarified through comprehensive transmission analysis, near-field distributions, and polarization-angle dependencies. The results demonstrate that the device supports two distinct resonant modes at the wavelength of around 789 nm and 814 nm, corresponding to $M_1$ and $M_2$, respectively, which together constitute a dual-mode AT mechanism. With x-polarized incidence, a single sharp resonance corresponding to $M_2$ dominates, where the forward transmissivity reaches 0.34 while the backward transmissivity remains below 0.05, yielding an isolation ratio of approximately 9.3 dB. In contrast, with y-polarized incidence, both $M_1$ and $M_2$ are supported, giving rise to a double-peak spectrum with the backward transmissivity less than 0.02 and the isolation ratio reaching 11 dB and

12.7 dB, respectively. These findings collectively verify that the AT arises from the robust interplay between LSPR coupling and grating-assisted decoupling, and the polarization properties arise from the orthometric gratings that behaves different varying with the polarization of the incident light. We develop the AO-GFG structure as a physically versatile platform for realizing high-isolation, polarization-tunable, and wavelength-selective plasmonic nanophotonic devices operating in the red and near-infrared regions.

## Acknowledgments

This work was supported by the National Natural Science Foundation of China (Grant No. 12504461).

## Disclosures

The authors declare no conflicts of interest.

## Data availability statement

The data that support the findings of this study are available upon reasonable request from the authors.

## References

1. C. Ba, L. Huang, W. Liu, *et al.*, "Narrow-band and high-contrast asymmetric transmission based on metal-metal-metal asymmetric gratings," Opt Express. **27**(18), 25107-25118 (2019).
2. A. E. Miroshnichenko, E. Brasselet and Y. S. Kivshar, "Reversible optical nonreciprocity in periodic structures with liquid crystals," Appl. Phys. Lett. **96**(6), 063302 (2010).
3. S. Cakmakyapan, H. Caglayan, A. E. Serebryannikov, *et al.*, "Experimental validation of strong directional selectivity in nonsymmetric metallic gratings with a subwavelength slit," Appl. Phys. Lett. **98**(5), 051103 (2011).
4. R. Zhu, X. Wu, Y. Hou, *et al.*, "Broadband Asymmetric Light Transmission at Metal/Dielectric Composite Grating," Sci Rep. **8**(1), 999 (2018).
5. E. Z. Li, D. S. Ding, Y. C. Yu, *et al.*, "Experimental demonstration of cavity-free optical isolators and optical circulators," Phys. Rev. Res. **2**(3), 033517 (2020).
6. H. Yang, C. Lou and X. Huang, "Broadband and highly efficient asymmetric optical transmission through periodic Si cylinder arrays on the dielectric substrates," Results Phys. **60**, 107691 (2024).

7. A. F. Popkov, M. Fehndrich, M. Lohmeyer, *et al.*, "Nonreciprocal TE-mode phase shift by domain walls in magnetooptic rib waveguides," Appl. Phys. Lett. **72**(20), 2508-2510 (1998).

8. Y. Shoji, T. Mizumoto, H. Yokoi, *et al.*, "Magneto-optical isolator with silicon waveguides fabricated by direct bonding," Appl. Phys. Lett. **92**(7), 071117 (2008).

9. X. N. Wu, G. W. Yuan, R. Zhu, *et al.*, "Giant Broadband One Way Transmission Based on Directional Mie Scattering and Asymmetric Grating Diffraction Effects," Chin. Phys. Lett. **37**(4), 044205 (2020).

10. Y. Shoji and T. Mizumoto, "Magneto-optical non-reciprocal devices in silicon photonics," Sci Technol Adv Mater. **15**(1), 014602 (2014).

11. J. A. Dionne, L. A. Sweatlock, H. A. Atwater, *et al.*, "Plasmon slot waveguides: Towards chip-scale propagation with subwavelength-scale localization - art. no. 035407," Phys. Rev. B. **73**(3), 035407 (2006).

12. V. A. Fedotov, P. L. Mladyonov, S. L. Prosvirnin, *et al.*, "Asymmetric propagation of electromagnetic waves through a planar chiral structure," Phys. Rev. Lett. **97**(16), 167401 (2006).

13. T. F. Gundogdu, M. Gokkavas, A. E. Serebryannikov, *et al.*, "Evidence of asymmetric beaming in a piecewise-linear propagation channel," Opt Lett. **46**(12), 2928-2931 (2021).

14. H. Liu, Y. Z. Zhang, C. Chen, *et al.*, "Dynamically adjustable and high-contrast asymmetric optical transmission based on bilateral compound metallic gratings," Opt Laser Technol. **140**, 107033 (2021).

15. D. Gao, W. Ding, M. Nieto-Vesperinas, *et al.*, "Optical manipulation from the microscale to the nanoscale: fundamentals, advances and prospects," Light Sci Appl. **6**(9), e17039 (2017).

16. M. Stolarek, D. Yavorskiy, R. Kotynski, *et al.*, "Asymmetric transmission of terahertz radiation through a double grating," Opt Lett. **38**(6), 839-841 (2013).

17. Y. Ling, L. Huang, W. Hong, *et al.*, "Asymmetric optical transmission based on unidirectional excitation of surface plasmon polaritons in gradient metasurface," Opt Express. **25**(12), 13648-13658 (2017).

18. A. E. Serebryannikov, E. Ozbay and S. Nojima, "Asymmetric transmission of terahertz waves using polar dielectrics," Opt Express. **22**(3), 3075-3088 (2014).

19. R. Jones, R. E. Camley and R. Macêdo, "Controlling asymmetric transmission in layered natural hyperbolic crystals," Opt. Laser Technol. **161**, 109210 (2023).

20. M. Vanwolleghem, X. Checoury, W. Smigaj, *et al.*, "Unidirectional band gaps in uniformly magnetized two-dimensional magnetophotonic crystals," Phys. Rev. B. **80**(12), 121102 (2009).

21. A. H. Gevorgyan, "Chiral photonic crystals with an anisotropic defect layer: Oblique incidence," Opt. Commun. **281**(20), 5097-5103 (2008).

22. M. W. Feise, I. V. Shadrivov and Y. S. Kivshar, "Bistable diode action in left-handed periodic structures," Phys. Rev. E. **71**(3), 037602 (2005).

23. A. B. Khanikaev and M. J. Steel, "Low-symmetry magnetic photonic crystals

for nonreciprocal and unidirectional devices," Opt Express. **17**(7), 5265-5272 (2009).

24. K. Gallo, G. Assanto, K. R. Parameswaran, *et al.*, "All-optical diode in a periodically poled lithium niobate waveguide," Appl. Phys. Lett. **79**(3), 314-316 (2001).

25. X. Hu, Y. Zhang, X. Xu, *et al.*, "Nanoscale Surface Plasmon All-Optical Diode Based on Plasmonic Slot Waveguides," Plasmonics. **6**(4), 619-624 (2011).

26. B. Khalichi, A. Ghobadi, A. K. Osgouei, *et al.*, "Diode like high-contrast asymmetric transmission of linearly polarized waves based on plasmon-tunneling effect coupling to electromagnetic radiation modes," J Phys D Appl Phys. **54**(36), 365102 (2021).

27. P. Zhang, Q. Leng, Y. S. Kan, *et al.*, "Asymmetric transmission of linearly polarized waves based on chiral metamaterials," Opt. Commun. **517**, 128321 (2022).

28. N. Peng, X. Li and W. She, "Nonreciprocal optical transmission through a single conical air hole in an Ag film," Opt Express. **22**(14), 17546-17552 (2014).

29. Z. W. Li, L. R. Huang, K. Lu, *et al.*, "Continuous metasurface for high-performance anomalous reflection," Appl. Phys. Express. **7**(11), 112001 (2014).

30. Y. Zhao and A. Alù, "Manipulating light polarization with ultrathin plasmonic metasurfaces," Phys. Rev. B. **84**(20), 205428 (2011).

31. N. Yu, P. Genevet, M. A. Kats, *et al.*, "Light propagation with phase discontinuities: generalized laws of reflection and refraction," Sci. **334**(6054), 333-337 (2011).

32. W. T. Chen, K. Y. Yang, C. M. Wang, *et al.*, "High-efficiency broadband meta-hologram with polarization-controlled dual images," Nano Lett. **14**(1), 225-230 (2014).

33. J. Luan, L. Huang, Y. Ling, *et al.*, "Dual-wavelength multifunctional metadevices based on modularization design by using indium-tin-oxide," Sci Rep. **9**(1), 361 (2019).

34. E. Battal, T. A. Yogurt and A. K. Okyay, "Ultrahigh Contrast One-Way Optical Transmission Through a Subwavelength Slit," Plasmonics. **8**(2), 509-513 (2012).

35. Y. Cheng, K. Zhai, N. Zhu, *et al.*, "Optical non-reciprocity with multiple modes in the visible range based on a hybrid metallic nanowaveguide," J. Phys. D: Appl. Phys. **55**(19), 195102 (2022).

36. P. Xu, X. Lv, J. Chen, *et al.*, "Dichroic Optical Diode Transmission in Two Dislocated Parallel Metallic Gratings," Nanoscale Res Lett. **13**(1), 392 (2018).

37. L. Meng, D. Zhao, Z. Ruan, *et al.*, "Optimized grating as an ultra-narrow band absorber or plasmonic sensor," Opt Lett. **39**(5), 1137-1140 (2014).

38. H. Liu, Y. Zhang, C. Chen, *et al.*, "Transverse-Stress Compensated Methane Sensor Based on Long-Period Grating in Photonic Crystal Fiber," IEEE Access. **7**, 175522-175530 (2019).

39. P. B. Johnson and R. W. Christy, "Optical Constants of the Noble Metals," Phys. Rev. B. **6**(12), 4370-4379 (1972).